\documentclass[final,5p,times,twocolumn]{elsarticle}

\usepackage{graphicx}

\usepackage{amssymb}

\usepackage{amsmath}

\usepackage[normalem]{ulem}  
\usepackage[dvips]{color} 

\renewcommand\sout{\bgroup \color{red} \ULdepth=-.5ex \ULset}

\newcommand{\Comments}[1]{}

\newcommand{\nn}{\nonumber}
\newcommand{\Vq}{{\cal V}_{q}}
\newcommand{\Vp}{V^{+}}
\newcommand{\Vm}{V^{-}}
\newcommand{\wt}{{\omega_\tau}}
\newcommand{\Feff}{\mathcal{F}_\mathrm{eff}}
\newcommand{\Seff}[1]{\mathcal{S}_\mathrm{eff}^\mathrm{#1}}
\newcommand{\ZNc}{Z_{N_c}}
\newcommand{\Zchi}{Z_\chi}
\newcommand{\TQ}{T_\mathrm{Q}}
\newcommand{\SCLQCD}{P-SC-LQCD}

\journal{Physics Letters B}

\begin{document}

\begin{frontmatter}



\title{Strong-coupling lattice study for QCD phase diagram\\
including both chiral and deconfinement dynamics}


\author[miura]{Kohtaroh Miura}
\address[miura]{INFN Laboratori Nazionali di Frascati, 
	I-00044, Frascati (RM), Italy}

\author[nakano]{Takashi Z. Nakano}
\address[nakano]{Department of Physics, Faculty of Science, Kyoto University, 
	Kyoto 606-8502, Japan\\
        Yukawa Institute for Theoretical Physics, 
	Kyoto University, Kyoto 606-8502, Japan}

\author[ohnishi]{Akira Ohnishi}
\address[ohnishi]{Yukawa Institute for Theoretical Physics, 
        Kyoto University, Kyoto 606-8502, Japan}

\author[kawamoto]{Noboru Kawamoto}
\address[kawamoto]{Department of Physics, Faculty of Science, 
        Hokkaido University, Sapporo 060-0810, Hokkaido, Japan}

\begin{abstract}
We investigate the QCD phase diagram
by using the strong-coupling expansion of the lattice QCD
with one species of staggered fermion and the Polyakov loop effective action
at finite temperature ($T$)
and quark chemical potential ($\mu$).
We derive an analytic expression of effective potential
$\mathcal{F}_{\mathrm{eff}}$
including both the chiral ($U_{\chi}(1)$)
and the deconfinement ($Z_{N_c}$)
dynamics with finite coupling effects
in the mean-field approximation.
The Polyakov loop increasing rate ($d\ell_p/dT$)
is found to have two peaks as a function of $T$
for small quark masses.
One of them is the chiral-induced peak
associated with the rapid decrease of the chiral condensate.
The temperature of the other peak is almost independent
of the quark mass or chemical potential,
and this peak is interpreted as the $Z_{N_c}$-induced peak.
\end{abstract}

\begin{keyword}
Lattice QCD \sep Extreme QCD \sep Strong-coupling \sep Deconfinement \sep Chiral Symmetry

\end{keyword}

\end{frontmatter}




The phase diagram for chiral and deconfinement transitions
in Quantum Chromodynamics (QCD)
at finite temperature ($T$) and/or quark chemical potential ($\mu$)
is one of the most fascinating subjects
in the current high energy and nuclear physics.
To investigate properties of QCD at high $T$ is one of the physics goals
in the LHC-ALICE experiments.
At $\mu=0$,
first principle investigations
based on lattice Monte-Carlo (LQCD-MC) simulations
predict pseudo-critical temperature,
$T_c\simeq 145-195~\mathrm{MeV}$,
for the chiral phase transition~\cite{Borsanyi:2010bp}.
In the finite $\mu$ region,
the LQCD-MCs suffer from the
notorious sign problem of the quark determinant,
and have not provided reliable results
in the large chemical potential region, $\mu/T>1$.

The strong-coupling ($1/g^2$) expansion
in the lattice QCD (SC-LQCD)
has been successful since
the beginning of the lattice gauge theory,
and would provide an alternative lattice framework to study
the QCD phase diagram including finite $\mu$ region.
In pure Yang-Mills theory,
the string tension in the strong-coupling limit
gives the area law~\cite{Wilson:1974sk},
and the LQCD-MC~\cite{Creutz} smoothly connects the strong-coupling result~\cite{Munster:1980iv}
to the scaling region.
In the pure Yang-Mills theory at finite $T$,
we can describe the $Z_{N_c}$ deconfinement transition based on
the effective potential for the Polyakov-loop ($\ell_p$)
in the leading order of the strong-coupling expansion
with the Haar measure effects~\cite{V_Ploop},
and higher order corrections have been investigated recently~\cite{Langelage}.
For the SC-LQCD including fermions,
many theoretical knowledge
have been accumulated so far~\cite{Montvay,Kawamoto:1981hw,KlubergStern:1982bs,KlubergStern:1983dg,DKS,DHK,Faldt1986,Bilic:1991nv,MDP},
and the chiral phase transition in the $T-\mu$ plane
have been well investigated in
the strong-coupling limit~\cite{NishiFuku,Azcoiti:2003eb,Kawamoto:2005mq,deForcrand:2009dh}.
It is remarkable that the coupling of
the chiral condensate $\sigma$ and the Polyakov-loop $\ell_p$
was extracted in the strong-coupling limit~\cite{IK_GO,Fukushima:2003fm},
and led to the invention of 
the Nambu-Jona-Lasino model
with Polyakov-loops (PNJL model)~\cite{Fukushima:2003fw}.
Recently, the finite lattice couping ($\beta=2N_c/g^2$)
effects are incorporated, and is found to give rise
to modifications of quark mass and chemical potential.
The evolution of the chiral phase transition
with increasing $\beta$
has been interpreted via these modifications~\cite{NLO,Nakano:2009bf}.
As will be shown later, this development opens a possibility to investigate
the finite coupling evolution of $Z_{N_c}$ deconfinement dynamics
in addition to the chiral dynamics in the whole $T-\mu$ plane~\cite{Miura:2010xu}.

One of the interesting observations
in the LQCD-MC is that peak positions of
chiral and Polyakov loop susceptibilities ($\chi_{\sigma,\ell_p}$)
are close to each other, and the small separation of them
could be explained as a consequence of the broad analytic behavior
of the crossovers~\cite{Borsanyi:2010bp}.
It would be meaningful
to ask ourselves whether the peak of $\chi_{\ell_p}$
is induced by the chiral crossover.
Otherwise, does the $Z_{N_c}$ dynamics
(accidentally or inevitably) leads to the peak of $\chi_{\ell_p}$
near the chiral crossover? To shed light on this problem,
it is a good strategy to investigate the finite $\mu$ cases.
In models such as PNJL model combined
with the statistical model~\cite{Fukushima:2010is} 
or Polyakov-Quark Meson model
with the functional renormalization group evolution~\cite{Herbst:2010rf},
two transitions almost coincide.
In the PNJL with a certain fit parameter set,
the first-order chiral phase transition
with a small jump of a small value of $\ell_p$
can be realized in the low $T$ and large $\mu$
region~\cite{Fukushima:2008wg,Tuominen,McLerran:2008ua,Sakai:2010rp}.

In this Letter, we investigate the chiral and
$Z_{N_c}$ deconfinement dynamics
by using the SC-LQCD with the Polyakov loop effects,
abbreviated as P-SC-LQCD.
The \SCLQCD\ is directly based on the lattice QCD,
and it does not contain any additional parameters than those in QCD.
The lattice coupling $\beta=2N_c/g^2$
in the plaquette action is a unique parameter of the lattice QCD in the chiral limit.
To investigate the chiral and $\ZNc$ deconfinement dynamics simultaneously,
we consider the effective action with leading [${\cal O}(1/g^0)$]
and next-to-leading order [NLO, ${\cal O}(1/g^2)$] effects
of the strong coupling expansion in the fermionic sector,
and the leading order contributions to the Polyakov-loop
[$\mathcal{O}(1/g^{2N_\tau})$] in the pure gluonic sector.
The present framework aims at developing previous
SC-LQCD studies for the chiral dynamics
\cite{DKS,Faldt1986,NishiFuku,Kawamoto:2005mq,NLO,Nakano:2009bf}
to include the $\ZNc$ deconfinement dynamics,
and leads to an extended version
of the IK-GO model~\cite{IK_GO,Fukushima:2003fm}
to include finite beta effects for the quark sector.
This framework allows us
to investigate the  beta evolution of
the interplay between chiral and $\ZNc$ dynamics consistently.
This is the advantage of using the \SCLQCD\ over effective models.

We briefly overview the derivation of the effective potential in \SCLQCD.
Details are shown in our previous papers~\cite{NLO,PNNLO_T}.
We start from the lattice QCD partition function
with one species of staggered fermion ($\chi$)
with a current quark mass ($m_0$) in the lattice unit $a=1$,
\begin{align}
\mathcal{Z}_{\mathrm{LQCD}}&= \int\mathcal{D}[\chi,\bar{\chi},U_{\nu}]
\exp\biggl[
-S_F-S_G-m_0\sum_x\bar{\chi}_x\chi_x
\biggr]\ ,\label{eq:Z}\\
S_F&= \frac{1}{2}\sum_{\nu,x}\Bigl[
\eta_{\nu,x}\bar{\chi}_x U_{\nu,x} \chi_{x+\hat{\nu}}
-\eta_{\nu,x}^{-1}(h.c.)
\Bigr]\ ,\label{eq:SF}\\
S_G&=\beta\sum_{P}
\biggl[
1-\frac{1}{2N_c}\Bigl[U_P+U_P^{\dagger}\Bigr]
\biggr]\label{eq:SG}\ .
\end{align}
where
$U_{\nu,x}\in SU(N_c)$
and
$U_{P=\mu\nu,x}=\mathrm{tr}_c[U_{\mu,x}U_{\nu,x+\hat{\mu}}U^\dagger_{\mu,x+\hat{\nu}}U^\dagger_{\nu,x}]$
represent the link-variable and plaquette, 
and 
the staggered sign factor
$\eta_{\nu,x}=\exp(\mu\,\delta_{\nu 0})(-1)^{x_0+\cdots +x_{\nu-1}}$
contains the the lattice chemical potential $\mu$.
The transformation $\chi_x\to e^{i\theta\epsilon_x}\chi_x$
with $\epsilon_x=(-1)^{x_0+\cdots+x_{d}}$
turns out to be a $U(1)_{\chi}$ chiral
transformation~\cite{Kawamoto:1981hw,KlubergStern:1983dg},
which leaves the staggered action $S_F$ invariant
in the chiral limit ($m_0\to 0$).
The $U_{\chi}(1)$ chiral symmetry
would be enhanced to $SU(N_f=4)$
in the continuum limit
\cite{KlubergStern:1983dg,Golt_Smit,stagg}.
The plaquette action $S_G$ is invariant under the global transformation
$U_{\nu,x}\to \Omega U_{\nu,x}$,
where $\Omega$ is the element of
the center of the $SU(N_c)$ gauge group, $Z_{N_c}$.
The chiral condensate
($\sigma \sim \langle\bar{\chi}\chi\rangle$)
and the Polyakov-loop
($L_p=\mathrm{tr}_c(\prod_{\tau}U_{0,\mathbf{x}\tau})/N_c$)
are the order parameters of the chiral and $\ZNc$ symmetries, respectively,
and they take finite values via the spontaneous breaking of these symmetries.
Both chiral and $Z_{N_c}$ symmetries
are explicitly broken for a finite quark mass $m_0$,
and the transitions could be replaced with the crossovers.
We concentrate on the color SU($N_c=3$) in the $3+1$ dimension ($d=3$)
in the later discussion.

\begin{figure}[bt]
\includegraphics[width=8.0cm]{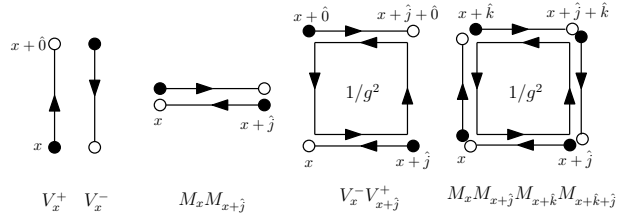}
\caption{Effective action terms
in the strong-coupling limit and $1/g^2$ corrections.
Open circles, Filled circles, and arrows show $\chi$, $\bar{\chi}$, and
$U_\nu$, respectively.}
\label{Fig:diagrams}
\end{figure}

The effective action of hadronic composites is obtained
by the Taylor expansion in $\beta$ and integrating out spatial link variables
in the finite $T$ treatment of \SCLQCD.
We include the leading and NLO terms in the fermionic sector,
$\Seff{NLO}$~\cite{NLO},
and we adopt the leading order Polyakov-loop effective action
$\Seff{Pol}$ in the pure gluonic sector~\cite{Kogut:1981ez},
\begin{align}
&\mathcal{Z}_{\mathrm{LQCD}}\simeq
\int\mathcal{D}[\chi,\bar{\chi},U_0]~
\exp\left(-\Seff{NLO}-\Seff{Pol}\right)
\ ,\\
&\Seff{NLO}
= \sum_x\Biggl[
\frac12 \left[ \Vp_x(\mu)-\Vm_x(\mu) \right]
+ m_0M_x\nn\\
&~~~+
\sum_{j>0}\biggl[
-\frac{M_xM_{x+\hat{j}}}{4N_c}
+\frac{\beta_{\tau}}{4d}
\Bigl[
\Vp_x(\mu) \Vm_{x+\hat{j}}(\mu)
+\Vp_x(\mu) \Vm_{x-\hat{j}}(\mu)
\Bigl]
\biggr]\nn\\
&~~~-
\frac{\beta_s}{d(d-1)}
		\sum_{0<k<j} 
		M_{x}
		M_{x+\hat{j}}
		M_{x+\hat{k}}
		M_{x+\hat{k}+\hat{j}}
\Biggr]
+\mathcal{O}\biggl(\frac{1}{g^4},\frac{1}{\sqrt{d}}\biggr)
\ ,\label{eq:Seff}
\\
&\Seff{Pol}[L_p,\bar{L}_p]
=-N_c^2\biggl(\frac{1}{g^2N_c}\biggr)^{N_{\tau}=1/T}
\sum_{j,\mathbf{x}}\Bigl[
\bar{L}_{p,\mathbf{x}}L_{p,\mathbf{x}+\hat{j}}+h.c.
\Bigr]\ ,\label{Eq:SeffP}
\end{align}
where the coupling $\beta_{\tau,s}$
and composites are defined as
$(\beta_{\tau},~\beta_s)=(\beta d/2N_c^3,~\beta d(d-1)/16N_c^5)$,
$M_x=\bar{\chi}_x\chi_x$,
and
$(\Vp_x(\mu),~\Vm_x(\mu))=
(e^{\mu}\bar{\chi}_xU_{0,x}\chi_{x+\hat{0}},~
e^{-\mu}\bar{\chi}_{x+\hat{0}}U_{0,x}^{\dagger}\chi_{x})$.
We consider only the leading order terms
in the $1/d$ expansion~\cite{KlubergStern:1982bs},
which corresponds
to the minimum quark number diagrams
for a given plaquette configuration
as shown in Fig.~\ref{Fig:diagrams}.
$\Seff{NLO}$ contains the four-Fermi interaction,
the temporal components of the vector interaction, and eight-Fermi interaction,
and $\Seff{Pol}$ shows
the nearest-neighbor interaction of the Polyakov loops.
%
We can reduce the effective action $\Seff{NLO}$
into the bi-linear form in staggered fermions
by introducing several auxiliary fields
($\sigma,~\omega_{\tau}$, and $\varphi_{\tau,s}$)
summarized in Table~\ref{Tab:aux},
\begin{align}
\Seff{NLO}
\simeq&
Z_\chi
\sum_{xy}\bar{\chi}_x G_{xy}^{-1}(\tilde{m}_q,\tilde{\mu})\chi_y\nn\\
+N_{\tau}N_s^d &
\Biggl[
\biggl(\frac{d}{4N_c}+\beta_s\varphi_s\biggr)\sigma^2
+\frac{\beta_s}{2}\varphi_s^2
+\frac{\beta_{\tau}}{2}
\bigl(
\varphi_{\tau}^2-\omega_{\tau}^2
\bigr)
\Biggr]\ ,
\label{Eq:SeffNLO}
\\
G_{xy}^{-1}(\tilde{m}_q,\tilde{\mu})
=&\tilde{m}_{q}\delta_{xy}
+\frac{\delta_{\mathbf{xy}}}{2}
\Bigl(
e^{\tilde{\mu}}U_{0,x}\delta_{x+\hat{0},y}
-e^{-\tilde{\mu}}U_{0,x}^{\dagger}\delta_{x-\hat{0},y}
\Bigr) \label{Eq:qhop}\ ,
\end{align}
where $N_{\tau}$ ($N_s$) represents the temporal (spatial) lattice size,
and all auxiliary fields are assumed to be constant and static.
The spontaneous breaking of the chiral symmetry with NLO effects results in
the dynamical shifts of quark mass
$\tilde{m}_{q}=\left[m_0+\left(d/2N_c+2\beta_s\varphi_s\right) \sigma
\right]/\Zchi$,
chemical potential
$\tilde\mu=\mu-\log\Zchi$,
and the quark wave function renormalization factor, $\Zchi$,
where 
$\Zchi=\sqrt{Z_+Z_-}$
and
$Z_{\pm}=1+\beta_{\tau}(\varphi_\tau\pm\omega_\tau)$.
Such a property allows us to utilize the technique developed
in the strong coupling limit,
and from a technical point of view,
it is essential to derive the effective potential
including both the chiral and deconfinement dynamics.

\begin{table}[hbt]
\caption{The auxiliary fields and their stationary values.
Here,
$\varphi_0=N_c-\Zchi\tilde{m}_q+\beta_{\tau}\omega_{\tau}^2$.
}\label{Tab:aux}
\begin{center}
\begin{tabular}{c|c|c}
\hline
Aux. Fields&Mean Fields &Stationary Values\\
\hline
$\sigma$        &$\langle -M\rangle$
                &$-\partial\Vq/\partial(\Zchi\tilde{m}_{q})$\\
$\varphi_s$     &$\langle MM\rangle$
                &$\sigma^2$\\
$\varphi_\tau$  &$-\langle (\Vp-\Vm)/2\rangle$
                &$2\varphi_0/(1+\sqrt{1+4\beta_\tau\varphi_0})$\\
$\wt$           &$-\langle (\Vp+\Vm)/2\rangle$
                &$-\partial\Vq/\partial\tilde{\mu}=\rho_q$\\
\hline
\end{tabular}
\end{center}
\end{table}
\begin{figure}[tbh]
\begin{center}
\includegraphics[width=8.0cm]{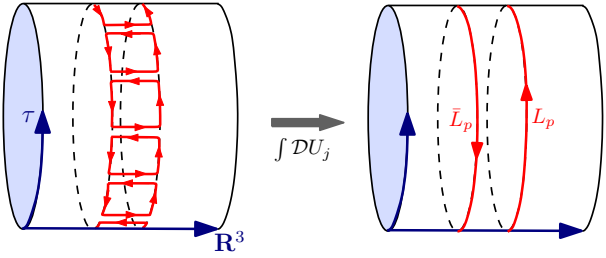}
\caption{Schematic figure of
the Polyakov-loop construction
in the strong-coupling expansion.
}\label{Fig:Ploop}
\end{center}
\end{figure}

Now we
perform the Gaussian integral
over the staggered quark fields $(\chi,\bar{\chi})$.
Here we work in the static and diagonalized gauge (called the Polyakov gauge)
for temporal link variables with respect for the periodicity~\cite{DKS},
\begin{align}
\mathcal{U}_{0,\mathbf{x}}
=\prod_{\tau}U_{0,\mathbf{x}\tau}=
\mathrm{diag}
\Bigl\{e^{i\theta_1(\mathbf{x})},\cdots,e^{i\theta_{N_c}(\mathbf{x})}\Bigr\}
\ .\label{eq:Pgauge}
\end{align}
Owing to the static property of the auxiliary fields 
and the temporal link variable in the Polyakov gauge,
the quark determinant is factorized in terms of the frequency modes,
and evaluated by the Matsubara method
(see for example the appendix 
in Ref.~\cite{Kawamoto:2005mq}).
The remarkable point is that
the resultant expression can be expressed
in terms of the Polyakov-loop variables ($L_p,\bar{L}_p$),
\begin{align}
&\int\mathcal{D}[\chi,\bar{\chi}]
\exp\biggl[
-\Zchi
\sum_{xy}\bar{\chi}_x G_{xy}^{-1}(\tilde{m}_q,\tilde{\mu})\chi_y
\biggr]\nn\\
&\quad
=
\Zchi^{N_cN_{\tau}N_s^d}
\prod_{\mathbf{x}}
\Bigl[
e^{N_cE_q/T}
\mathcal{R}_{q}(T,\mu)
\mathcal{R}_{\bar{q}}(T,\mu)
\Bigr]\ ,\label{Eq:ZqLp}\\
&\mathcal{R}_{q}(T,\mu)
\equiv
1+e^{-N_c(E_q-\tilde{\mu})/T}\nn\\
&\quad +N_c\Bigl(L_{p,\mathbf{x}}e^{-(E_q-\tilde{\mu})/T}
+\bar{L}_{p,\mathbf{x}}e^{-2(E_q-\tilde{\mu})/T}\Bigr)
\ ,\label{eq:Dq}\\
&\mathcal{R}_{\bar{q}}(T,\mu)
\equiv
1+e^{-N_c(E_q+\tilde{\mu})/T}\nn\\
&\quad +N_c\Bigl(\bar{L}_{p,\mathbf{x}}e^{-(E_q+\tilde{\mu})/T}
+L_{p,\mathbf{x}}e^{-2(E_q+\tilde{\mu})/T}\Bigr)
\ .\label{eq:Dqb}
\end{align}
where $E(\tilde{m}_{q}(\sigma))=\sinh^{-1}\bigl[\tilde{m}_{q}(\sigma)\bigr]$
corresponds to the quark excitation energy.
In that expression,
$L_p$ couples to a Boltzmann factor $e^{-(E_q-\tilde{\mu})/T}$,
and determines how quarks thermally excites.
In the confined phase ($L_{p} \sim 0$),
color-singlet states dominate,
while quarks can excite in the deconfined phase ($L_{p} \neq 0$).
Equations~(\ref{eq:Dq}) and (\ref{eq:Dqb})
give a natural coupling manner between $L_p$ and $\sigma$.
This point has been pointed out
in the strong-coupling limit~\cite{IK_GO,Fukushima:2003fm},
and utilized in the PNJL model~\cite{Fukushima:2003fw}.
In the current formulation,
the Boltzmann factor
includes the NLO effects in $(\tilde{m}_q,\tilde{\mu})$,
and the coupling manner between $L_p$ and $\sigma$
is modified by the finite coupling effects.

Finally, we evaluate the temporal link integral
and obtain the effective potential.
In the Polyakov gauge,
the Haar measure becomes a Van der Monde determinant over the color space,
and can be rewritten by using the (reduced) Polyakov-loop
($L_{p,\mathbf{x}}\to \sum_{a=1}^{N_c}e^{i\theta_a(\mathbf{x})}/N_c$),
\begin{align}
&\int d\mathcal{U}_{0,\mathbf{x}}
=27
\int d[L_{p,\mathbf{x}},\bar{L}_{p,\mathbf{x}}]
\mathcal{M}_{\mathrm{Haar}}(L_p,\bar{L}_p)\ ,\\
&\mathcal{M}_{\mathrm{Haar}}(L_p,\bar{L}_p)
=1-6\bar{L}_{p,\mathbf{x}}L_{p,\mathbf{x}}
-3\bigl(\bar{L}_{p,\mathbf{x}}L_{p,\mathbf{x}}\bigr)^2
+4\bigl(L_{p,\mathbf{x}}^{N_c}+\bar{L}_{p,\mathbf{x}}^{N_c}\bigr)
\ .
\label{Eq:LpHaar}
\end{align}
While it is possible to perform this integral exactly,
we here adopt a simpler prescription;
we replace Polyakov-loops in the integrand
with its constant mean-field value,
$(L_{p,\mathbf{x}},\bar{L}_{p,\mathbf{x}})\to (\ell_{p},\bar{\ell}_{p})$,
and search for the stationary values of
$(\ell_{p},\bar{\ell}_{p})$.
This treatment gives the effective potential
in a similar expression to that used in the PNJL model,
and useful for the comparison.
Thus, we obtain the effective potential
as a function of the auxiliary fields
$\Phi=(\sigma,\varphi_{\tau,s},\omega_{\tau},\ell_p,\bar{\ell}_p)$,
temperature $T$, and 
quark chemical potential $\mu$ in the mean-field approximation,
\begin{align}
&\Feff(\Phi;T,\mu)
\equiv -(T\log \mathcal{Z}_{\mathrm{LQCD}})/N_s^d
=\Feff^{\chi}+\Feff^\mathrm{Pol}
\ ,\label{eq:Feff}
\\
&\Feff^\chi
\simeq
\biggl(
\frac{d}{4N_c}+\beta_s\varphi_s
\biggr)\sigma^2
+\frac{\beta_s\varphi_s^2}{2}
+\frac{\beta_{\tau}}{2}
\bigl(\varphi_{\tau}^2-\omega_{\tau}^2\bigr)
-N_c\log \Zchi
\ ,
\nn\\
&~~~~~~~~~~
-N_cE_q
-T\bigl(\log\mathcal{R}_{q}(T,\mu)+\log\mathcal{R}_{\bar{q}}(T,\mu)\bigr)
\ ,\label{eq:FeffChi}
\\
&\Feff^\mathrm{Pol}\simeq
-2TdN_c^2\biggl(\frac{1}{g^2N_c}\biggr)^{1/T}\bar{\ell}_{p}\ell_{p}
-T\log\mathcal{M}_{\mathrm{Haar}}(\ell_p,\bar{\ell}_p)
\ ,\label{eq:FeffPol}
\end{align}
The equilibrium is determined
by imposing stationary conditions
on the effective potential, 
$\partial{\Feff}/\partial\Phi=0$,
which lead to the relations summarized
in the third column of Table \ref{Tab:aux}.
$\Feff^\chi$ is responsible for the chiral-dynamics,
and $\Feff^\mathrm{Pol}$ originates from the
plaquette action and governs the $Z_{N_c}$ dynamics.
These ingredients communicate with each other
through the quark determinant effects
{\em i.e.} $\mathcal{R}_{q}(T,\mu)$
and $\mathcal{R}_{\bar{q}}(T,\mu)$
in Eq.~(\ref{eq:FeffChi}).

The $\bar{\ell}_{p}\ell_{p}$ term 
in Eq.~(\ref{eq:FeffPol})
gives large finite $T$ effects
to $Z_{N_c}$ dynamics at finite $\beta$,
and vanishes in the strong-coupling limit.
In the previous work in the strong coupling limit,
this quadratic term is fixed to a constant
to be consistent with the empirical value of the string tension
\cite{Fukushima:2003fm}.
The finite coupling property of the current formulation
allows us to investigate the $\beta$ evolution
of the $\bar{\ell}_{p}\ell_{p}$ term, {\em i.e.}
finite $T$ effects of $Z_{N_c}$ dynamics,
without introducing additional parameters.

\begin{figure}[ht]
\begin{center}
\includegraphics[width=7.0cm]{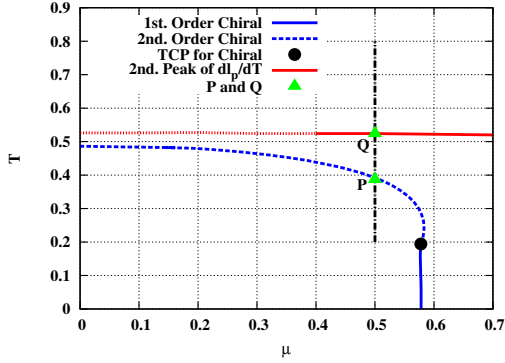}
\caption{
The phase boundary for the chiral transition
with the second peak of $d\ell_p/dT$
at $\beta=4$ in the chiral limit.
The ``P'' and ``Q'' correspond to
those in Fig.~\protect\ref{Fig:OrdEq}
and in the upper panel of Fig.~\protect\ref{Fig:DLp}.}
\label{Fig:PD}
\end{center}
\end{figure}

\begin{figure}[ht]
\begin{center}
\includegraphics[width=7.0cm]{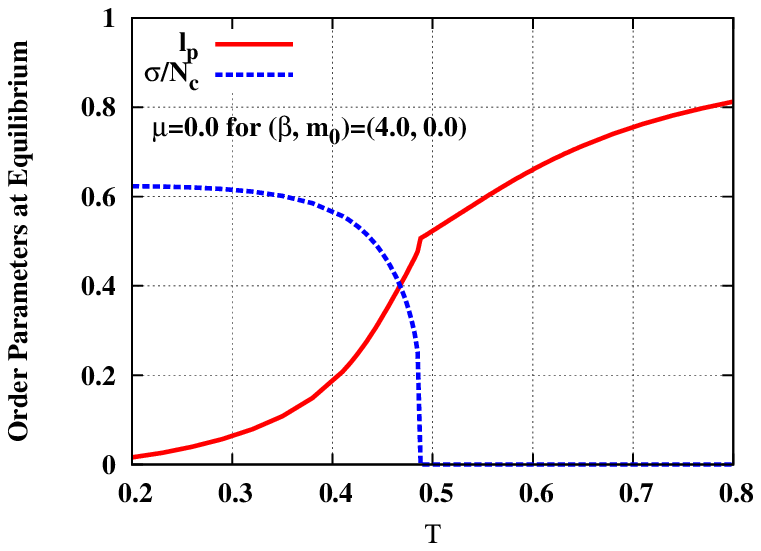}
\includegraphics[width=7.0cm]{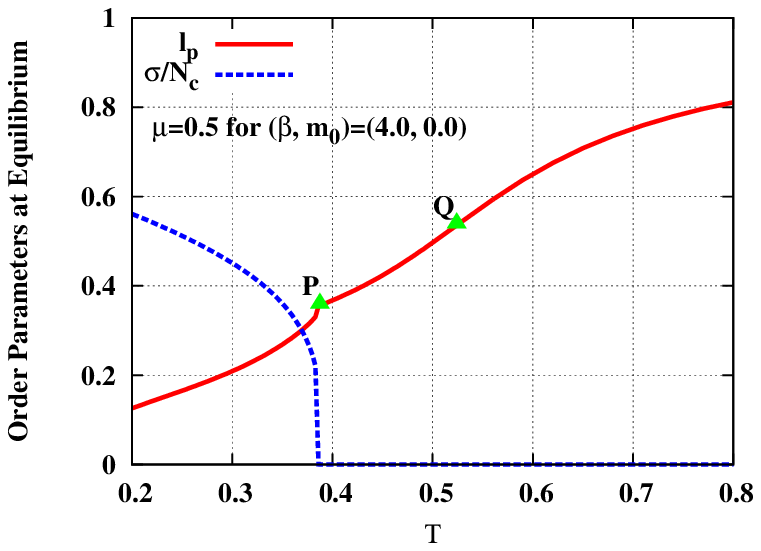}
\caption{
The $T$ dependence of the Polyakov loop
and the chiral condensate
for $\mu=0$ (upper) and $0.5$ (lower)
at $(\beta,m_0)=(4,0)$.
In the lower panel,
the ``P'' and ``Q'' correspond to
those in Fig.~\protect\ref{Fig:PD}
and the upper panel of Fig.~\protect\ref{Fig:DLp}.}
\label{Fig:OrdEq}
\end{center}
\end{figure}

\medskip

We shall show the first \SCLQCD\ results including
the ``chiral and deconfinement'' dynamics,
``finite $\mu$ effects'', and ``finite coupling effects'' simultaneously.
As shown in our recent work~\cite{PNNLO_T},
the critical temperature at zero chemical potential
becomes closer to the LQCD-MC results 
in the coupling region $\beta \lesssim 4$.
Therefore, we focus our attention to the results at
$\beta=4$. In the last part,
we discuss the stability of our main conclusion
for variations of $\beta$.

In Fig.~\ref{Fig:PD},
the phase diagram
at $(\beta,m_0)=(4,0)$
is shown in the lattice unit.
We find the first- (solid blue)
and second-order (dashed blue)
chiral transition lines separated by
the (tri-)critical point (CP)
at $(\mu_{\mathrm{CP}},T_{\mathrm{CP}})=(0.58,0.19)$.
The result is qualitatively
consistent with the previous SC-LQCD
with NLO effects~\cite{NLO}.
In Fig.~\ref{Fig:OrdEq},
we show the $T$ dependence
of the chiral condensate $\sigma$
and the Polyakov loop $\ell_p$.
The upper (lower) panel displays
the results for $\mu=0~(0.5)$,
{\em i.e.~}on the $T$ axis (dash-dotted line) of
the phase diagram in Fig.~\ref{Fig:PD}.
We note $\mu=0.5<\mu_{\mathrm{CP}}$,
and the following results would not be contaminated
by the fluctuation around the CP.

\begin{figure}[ht]
\begin{center}
\includegraphics[width=7.0cm]{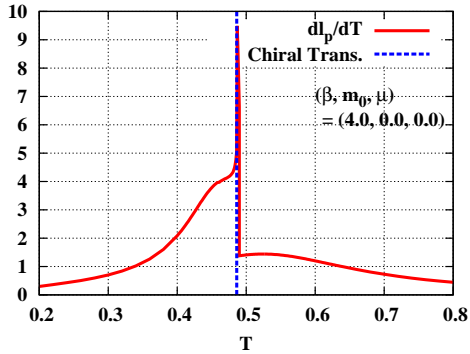}
\caption{The $T$ dependence of
$d\ell_p/dT$ in the chiral limit
with zero chemical potential for $\beta=4$.
The values are shown in the lattice unit $a=1$.
The dashed blue line
represents the critical temperature $T_{c,\mu=0}$
for the second-order chiral transition.}
\label{Fig:DLp00}
\end{center}
\end{figure}

We find two peaks in $d \ell_p/dT$.
One peak appears at the chiral phase transition.
For both $\mu=0$ and $0.5$ cases,
the Polyakov loop $\ell_p$ shows a rapid increase
at the chiral transition temperature ($\sigma\to 0$).
The strong correlation of the Polyakov loop 
{and $\sigma$}
is found {at any point on the chiral transition boundary}.
This correlation can be seen more clearly
in the derivative $d\ell_p/dT$ as shown in Fig.~\ref{Fig:DLp00}.
We find a sharp peak in the vicinity of
the chiral phase transition,
$T_{c,\mu=0}(\beta=4)\simeq 0.486$.
The almost simultaneous observation
of the chiral transition and the rapid change of $\ell_p$
would be consistent with the LQCD-MC results~\cite{Borsanyi:2010bp}.
Via the $\sigma$-$\ell_p$ coupling in the Boltzmann factor terms
in Eq.~(\ref{eq:Dq}) and (\ref{eq:Dqb}),
the chiral transition would induce the rapid change of $\ell_p$.
Thus we regard this peak as the chiral-induced peak.

\begin{figure}[ht]
\begin{center}
\includegraphics[width=7.0cm]{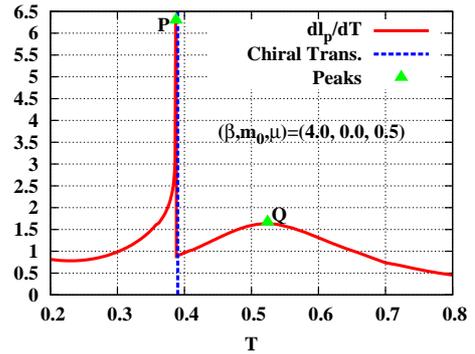}
\includegraphics[width=7.0cm]{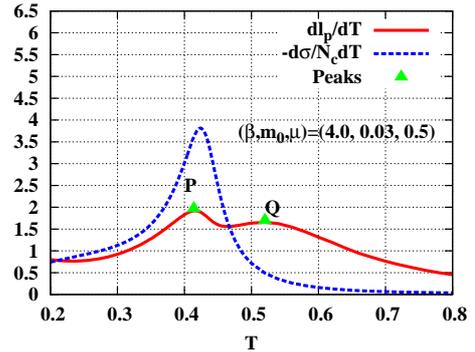}
\includegraphics[width=7.0cm]{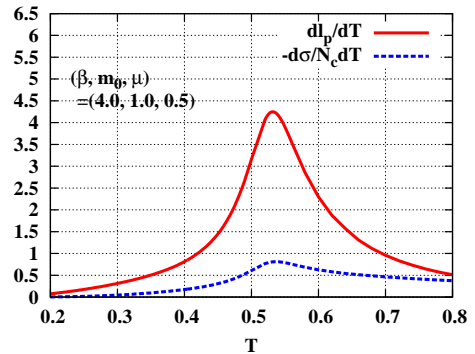}
\caption{
$T$ dependence of $d\ell_p/dT$
at $\mu=0.5$ for $\beta=4$ in the lattice unit.
The current quark masses ($m_0$) are
$0.0$ (upper), $0.03$ (middle), and $1$ (bottom).
The dashed lines in the middle and lower panels represent
$-(N_c)^{-1}d\sigma/dT$.
\label{Fig:DLp}}
\end{center}
\end{figure}

We also find the second peak above the chiral transition 
temperature in $d\ell_p/dT$.
At $\mu=0$, a small enhancement is seen at around $T \simeq 0.52$.
At finite $\mu$, 
the enhancement of $d\ell_p/dT$ at the second peak becomes significant.
A similar double-peak structure has been reported
in the model studies based on PNJL model~\cite{Tuominen}.
In Fig.~\ref{Fig:DLp}, we show $d\ell_p/dT$
along the line with $\mu=0.5$ (dash-dotted line in Fig.~\ref{Fig:PD}).
In the chiral limit ($m_0\to 0$, upper panel),
we find two peaks ``P'' and ``Q''.
Here, ``P'' and ``Q'' in Fig.~\ref{Fig:DLp}
correspond to those in Figs.~\ref{Fig:PD} and \ref{Fig:OrdEq}.
The peak ``P'' locates at the chiral phase transition,
and is clearly interpreted as the chiral-induced peak.
As indicated by the red line in Fig.~\ref{Fig:PD},
the similar peak to ``Q'' is observed in the whole range of $\mu$.
The strength of this peak becomes weaker for smaller $\mu$,
which is expressed by the dotted line in Fig.~\ref{Fig:PD}.

The peak ``Q'' can be understood as a signal
of the $\ZNc$-induced crossover:
$\ZNc$ is the symmetry in the pure gluonic sector,
and it becomes exact in the heavy quark mass limit.
We also expect weak $\mu$ dependence of the $\ZNc$ deconfinement transition,
since $\Feff^\mathrm{Pol}$ does not directly depend
on quark chemical potential $\mu$.
In the middle and lower panels of Fig.~\ref{Fig:DLp},
we show $d\ell_p/dT$ for $m_0=0.03$ and $m_0=1$, respectively.
For $m_0=0.03$, we find two peaks.
The temperature of the first peak ``P''
is slightly shifted upward and becomes closer to ``Q'' ($\TQ$)
with increasing $m_0$,
while $\TQ$ stays almost constant.
For larger masses, $m_0>0.05$,
the two peaks merges to a single peak,
as shown in the lower panel of Fig.~\ref{Fig:DLp} for $m_0=1$.
This single peak grows with increasing $m_0$,
and its temperature is nearly $m_0$ independent, $T\sim0.52$,
which is close to $\TQ$ at smaller quark masses.
The $\ZNc$-induced nature of ``Q'' and the merged single peak are confirmed by
the weak dependence of $\TQ$ on $\mu$ and $m_0$
as found in Figs.~\ref{Fig:PD} and \ref{Fig:DLp}, respectively.
It is interesting to find that 
the $Z_{N_c}$ nature survives in the chiral limit or small mass region,
and can be observed as a peak in $d\ell_p/dT$.

{There are several comments in order.}
(a) In cold dense matter in Fig.~\protect\ref{Fig:PD},
the chiral symmetry is restored and Polyakov loop is suppressed.
These features may be similar
to those of quarkyonic matter~\cite{McLerran:2008ua}.
(b) We find that $\TQ$ is larger than the chiral phase transition temperature
for small quark masses. In the chiral limit,
the Polyakov loop $\ell_p$ decouples from the chiral dynamics
at $T=\TQ$ since $\sigma$-$\ell_p$ couplings in Eq.~(\ref{eq:Feff})
vanish due to the exact chiral restoration $\sigma=0$.
(c) We note that
the chiral condensate decreasing rate, $-d\sigma/dT$,
has a small peak at the $Z_{N_c}$-deconfinement crossover
for $m_0=1$ as shown in the lower panel of Fig.~\ref{Fig:DLp}.
This peak is interpreted as a $Z_{N_c}$-induced peak.
In this case, however,
another peak which would stem from
the chiral symmetry is completely
overwhelmed due to the large quark mass,
and the double-peak structure does not appear.

\begin{figure}[ht]
\begin{center}
\includegraphics[width=7.0cm]{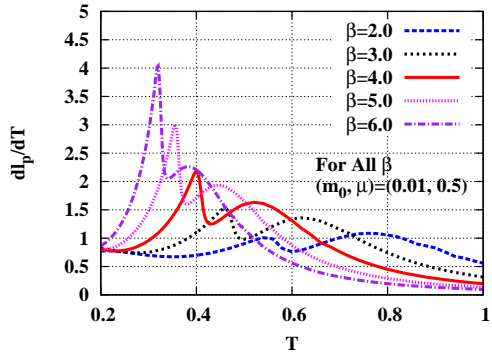}
\caption{The $T$ dependence of
$d\ell_p/dT$ for several $\beta$ at $(m_0,\mu)=(0.01,0.5)$.}
\label{Fig:DLp_beta}
\end{center}
\end{figure}

Finally, we discuss the $\beta$ dependence
of the two peak structure of $d\ell_p/dT$
at finite $\mu$.
In Fig.~\ref{Fig:DLp_beta},
we show $d\ell_p/dT$ for several $\beta$ at $(m_0,\mu)=(0.01,0.5)$.
As indicated from this figure,
the two peaks are found at least in the range
$2\le \beta\le 6$.
For each $\beta$,
we define the pseudo-critical temperatures
for the chiral-induced ($T_{c,\mu=0.5}^{(\chi)}(\beta)$)
and $Z_{N_c}$-induced ($T_{c,\mu=0.5}^{(d)}(\beta)$)
deconfinement crossovers
as the first and second peaks of $d\ell_p/dT$, respectively.
Both of them are decreasing functions of $\beta$.
For $T_{c,\mu=0.5}^{(\chi)}(\beta)$,
the current results are close to
the pseudo-critical temperatures for
the chiral phase transition in
our previous works~\cite{NLO,Nakano:2009bf}.
This again indicates the chiral induced nature of
the first peak ``P''.

We find that the $\beta$ dependence of
$T_{c,\mu=0.5}^{(d)}(\beta)$ is larger
than that of $T_{c,\mu=0.5}^{(\chi)}(\beta)$,
and the separation between two peaks
tends to be narrower with increasing $\beta$
when chemical potential is fixed, $\mu=0.5$.
Whereas the two peaks tend to be more separated
for larger chemical potential for a fixed value of $\beta$.
For example, two peak separation at $\mu=0.55$
becomes $1.4$ times larger than that at $\mu=0.5$ for $\beta=6$.
This is because the chiral-induced transition temperature decreases
in the large $\mu$ region of the $T-\mu$ plane,
while there is no direct $\mu$ dependence in the $Z_{N_c}$ dynamics.

\medskip
In summary,
we have investigated the chiral and deconfinement crossovers
at finite temperature $T$ and quark chemical potential $\mu$
based on the strong-coupling expansion in the lattice QCD
with one species of staggered fermion.
We have considered the leading and NLO effects
in the strong-coupling expansion for fermionic sector,
and the leading order Polyakov-loop effective action terms
for the pure Yang-Mills sector.
The $Z_{N_c}$ deconfinement dynamics
has been incorporated to the Haar measure,
where the Polyakov-loop is replaced with its constant mean-field value.

We have found double-peak structure
in the Polyakov loop increasing rate $d\ell_p/dT$
as a function of $T$
for small quark masses and large $\mu$.
The first peak is induced by the chiral transition.
This is because the Polyakov loop $\ell_p$
becomes sensitive to the chiral dynamics
through the coupling to the chiral condensate $\sigma$.
For the larger quark mass $m_0$,
the first peak is overwhelmed by the second peak,
whose position is almost independent of $m_0$.
This indicates that
the second peak would attribute to
the remnant $Z_{N_c}$ dynamics,
which is less affected by $\mu$ than the chiral phase transition line
due to the lack of a direct $\mu$ dependence
in the Haar measure treatment for $Z_{N_c}$.
Hence the double peaks, {\em i.e.~}
the chiral and $Z_{N_c}$ induced peaks, come out
in large $\mu$ region.

As future perspectives, we {should} evaluate
the chiral and Polyakov-loop susceptibilities,
next-to-next-to-leading order (NNLO) terms
of strong coupling expansion for the chiral dynamics~\cite{Nakano:2009bf}, 
higher order corrections of the Polyakov-loop effective action~\cite{Langelage},
and higher order terms of the $1/d$ expansion.
The exact evaluation in each order of
the strong-coupling expansion is also expected
by the Monomer-Dimer-Polymer formulation \cite{MDP,deForcrand:2009dh}.
These corrections include couplings between
the Polyakov loop and the chiral sector
in the effective action level.
Taking account of the resultant entanglement effects of
the chiral and $Z_{N_c}$ dynamics,
the appearance of chiral and $Z_{N_c}$ induced peaks
must be investigated in future.

We would like to thank Maria Paola Lombardo, Lars Zeidlewicz,
Philippe de Forcrand, Michael Fromm, and Kim Splittorff
for fruitful discussions.
We also thank Zoltan Fodor for useful comments for
the critical temperature.
This work was supported in part
by Grants-in-Aid for Scientific Research from JSPS
(No. 22-3314),
the Yukawa International Program for Quark-hadron Sciences (YIPQS),
and by Grants-in-Aid for the global COE program
``The Next Generation of Physics, Spun from Universality and Emergence''
from MEXT.





\end{document}